%% file: motion_unblocking.tex
\documentclass[acmtog]{acmart}
\acmSubmissionID{1425}

\usepackage{booktabs} 
\usepackage{multirow}
\usepackage{listings} 
\usepackage{caption}
\usepackage{xcolor}
\usepackage{soul}

\definecolor{codegreen}{rgb}{0,0.6,0}
\definecolor{codegray}{rgb}{0.5,0.5,0.5}
\definecolor{codepurple}{rgb}{0.58,0,0.82}
\definecolor{backcolour}{rgb}{0.95,0.95,0.92}

\lstdefinelanguage{mypython}{
  language=Python,
  morekeywords={self, None, True, False},
  sensitive=true,
  morecomment=[l]\#,
  morestring=[b]',
  morestring=[b]"
}

\lstdefinestyle{mystyle}{
  language=mypython,
  backgroundcolor=\color{backcolour},
  commentstyle=\color{codegreen}\itshape,
  keywordstyle=\color{magenta}\bfseries,
  numberstyle=\tiny\color{codegray},
  stringstyle=\color{codepurple},
  emph={loose_infill, track, attribute_select}, 
  emphstyle=\color{blue},
  basicstyle=\ttfamily\footnotesize,
  breaklines=true,
  breakautoindent=false,
  frame=single,
  rulecolor=\color{backcolour},
  captionpos=b,
  numbers=left,
  numbersep=5pt,
  showstringspaces=false,
  tabsize=2
}

\usepackage[ruled]{algorithm2e} 

\SetAlFnt{\small}
\SetAlCapFnt{\small}
\SetAlCapNameFnt{\small}
\SetAlCapHSkip{0pt}

\input{customMacros}

\acmJournal{TOG}

\copyrightyear{2025}
\acmYear{2025}
\setcopyright{acmlicensed}\acmConference[SA Conference Papers '25]{SIGGRAPH Asia 2025 Conference Papers}{December 15--18, 2025}{Hong Kong, Hong Kong}
\acmBooktitle{SIGGRAPH Asia 2025 Conference Papers (SA Conference Papers '25), December 15--18, 2025, Hong Kong, Hong Kong}
\acmDOI{10.1145/3757377.3763874}
\acmISBN{979-8-4007-2137-3/2025/12}

\begin{document}
\title{Generating Detailed Character Motion from Blocking Poses}


\author{Purvi Goel}
\orcid{1234-5678-9012-3456}
\affiliation{%
 \institution{Stanford University}
 \country{USA}
}
\email{pgoel2@cs.stanford.edu}
\author{Guy Tevet}
\affiliation{%
 \institution{Stanford University}
 \country{USA}
}
\email{jwang23@snapchat.com}
\author{C. Karen Liu}
\affiliation{%
 \institution{Stanford University}
 \country{USA}
}
\email{karenliu@cs.stanford.edu}
\author{Kayvon Fatahalian}
\affiliation{%
 \institution{Stanford University}
 \country{USA}
}
\email{kayvonf@cs.stanford.edu}


\begin{abstract}
\input{abstract}
\end{abstract}

\begin{teaserfigure}
\centerline{
\includegraphics[page=1,trim={10 0 0 0},clip,width=0.9\linewidth]{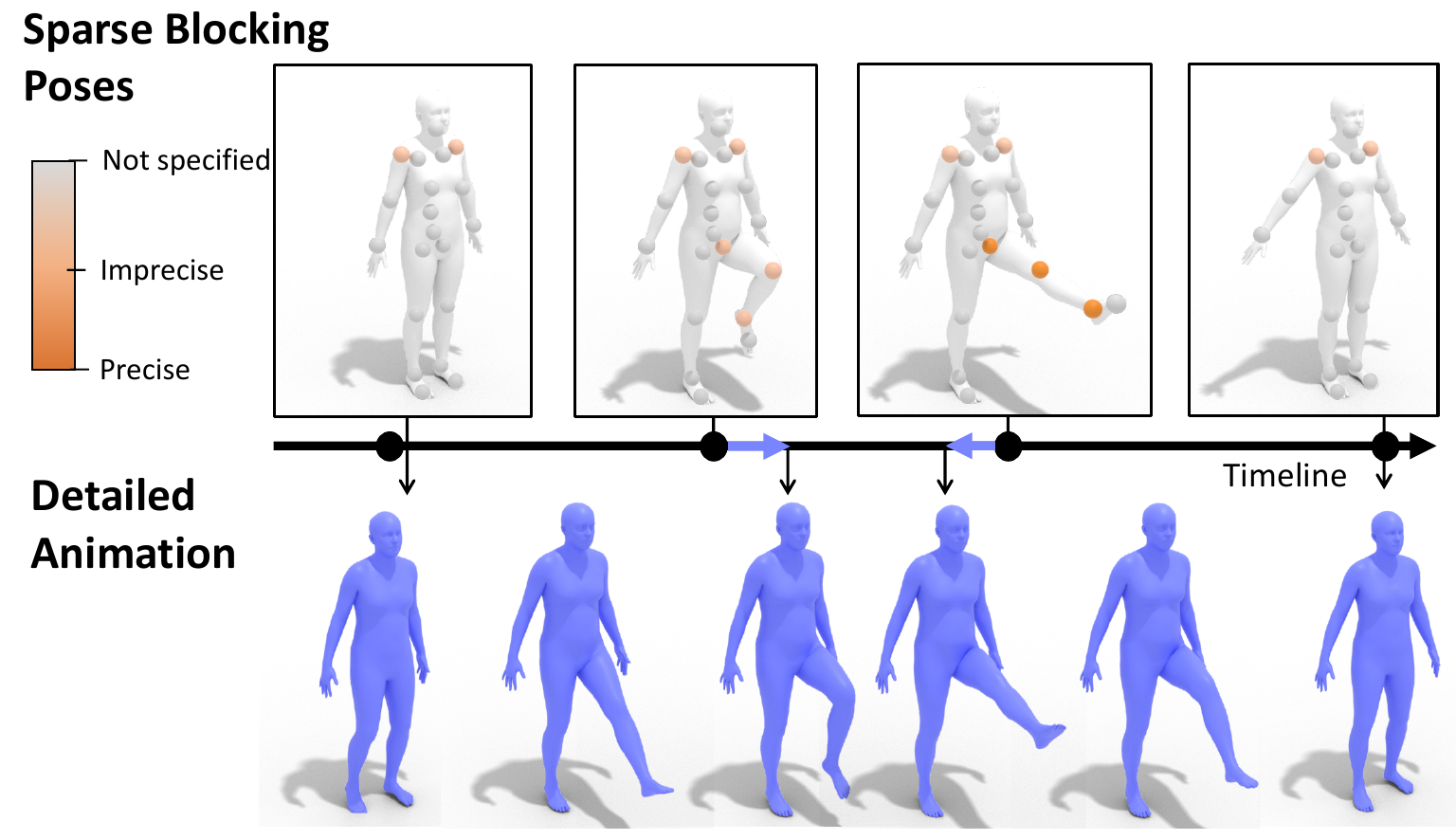}
}
\caption{Our method supports motion detailing of ``blocking poses'' (top), sparse sets of representative poses created by animators early in the animation process that give the gist of the timing and articulation of a character's motion, but may be imprecise or incomplete. Given these blocking poses and user-provided tolerance weights (orange), our method simultaneously refines the blocking poses according to their tolerance weights, retime them when necessary (blue arrows), and generates a natural-looking motion sequence (bottom) with detailed transitions, such as steps, a snappy kick, and arm movements for balance.
}
\label{fig:banner}
\end{teaserfigure}

\maketitle

\input{introduction}

\input{relatedwork}

\input{method}

\input{results}

\input{conclusion}

\bibliographystyle{ACM-Reference-Format}
\bibliography{motion_unblocking}

\end{document}

%% file: customMacros.tex
\usepackage{ifthen}

\newboolean{showComment}
\setboolean{showComment}{true} 

\ifthenelse{\boolean{showComment}}{
    
\newcommand{\purvi}[1]{\textcolor{blue}{[purvi]: #1}}
\newcommand{\needcite}[1]{\textcolor{red}{[CITE: #1}}
\newcommand{\karen}[1]{\textcolor{red}{[Karen: #1]}}

\newcommand{\revision}[1]{\textcolor{black}{#1}}
}{

\newcommand{\purvi}[1]{}
\newcommand{\KF}[1]{}
\newcommand{\needcite}[1]{}
\newcommand{\karen}[1]{}
\newcommand{\s}[1]{}
}

%% file: abstract.tex
We focus on the problem of using generative diffusion models for the task of motion detailing: converting a rough version of a character animation, represented by a sparse set of coarsely posed, and imprecisely timed blocking poses, into a detailed, natural looking character animation.  Current diffusion models can address the problem of correcting the timing of imprecisely timed  poses, but we find that no good solution exists for leveraging the diffusion prior to enhance a sparse set of blocking poses with additional pose detail. We overcome this challenge using a simple inference-time trick. At certain diffusion steps, we blend the outputs of an unconditioned diffusion model with input blocking pose constraints using per-blocking-pose tolerance weights, and pass this result in as the input condition to an pre-existing motion retiming model. We find this approach works significantly better than existing approaches that attempt to add detail by blending model outputs or via expressing blocking pose constraints as guidance. The result is the first diffusion model that can robustly convert blocking-level poses into plausible detailed character animations. 

%% file: introduction.tex
\section{Introduction}
\label{sec:intro}

Diffusion-based methods offer the promise of generating detailed character animations from high-level specifications such as text prompts~\cite{tevet2023human,athanasiou2024motionfixtextdriven3dhuman,ren2023insactor,zhang2022motiondiffusetextdrivenhumanmotion,zhang2023finemogen,wu2024humanobjectinteractionhumanlevelinstructions,karunratanakul2023guided}, trajectories~\cite{shafir2024human,Zhong:2025:Sketch2Anim,xie2023omnicontrol,rempeluo2023tracepace}, sparse keyframes~\cite{cohan2024flexible,goel2025generative,wei2023enhanced}, or even the music a character should dance to~\cite{tseng2022edge}. While these offer interesting new controls for animation synthesis, one of the most common workflow in professional character animation begins with \emph{motion blocking}: identifying poses (called ``blocking poses'' or ``keys'') that convey the desired action, and placing them on a timeline to establish a rough sketch of the timing, positioning, and posing of an envisioned motion. Blocking poses are typically few in number, imprecise in timing, and depict coarse estimates of pose (e.g., only a few joints are posed). Rather than precise specification, blocking poses are intended as scaffolding for future \emph{motion detailing} passes where an animator fills in the details to bring the motion to life, while adjusting the timing and posing of the blocking poses themselves as necessary ~\cite{animatorssurvivalkit, principlesofanimation3d, gameanim}. 

Unfortunately, there is no existing diffusion technique that robustly performs motion detailing: converting blocking poses to a plausible, detailed character animation.  Recently, \citet{goel2025generative} took a step in this direction by contributing a method for adjusting the timing of imprecisely timed, sparse keyframes. However, their method assumes detailed keyframe inputs, not blocking poses. As a result, running their method on input blocking poses results in animations exhibiting the same coarse level of pose detail as the blocking poses. We find that in this setting, standard ways for leveraging the diffusion prior to enhance pose detail, such as blending diffusion outputs with desired keyframes~\cite{shafir2024human, tseng2022edge}, or using keyposes as external constraints in reconstruction guidance~\cite{xie2023omnicontrol}, fail to produce plausible motions.

We show that a simple inference-time trick, used in conjunction with the motion retiming model of Goel et al.\,\shortcite{goel2025generative}, surprisingly meets our motion-detailing goals. Specifically, at certain diffusion steps, we blend the outputs of an unconditioned diffusion model with the input blocking poses based on animator-controlled \textit{per-blocking-pose} tolerance weights, and pass this blended result as an \text{input condition} to the existing retiming model. This approach contrasts with standard blending approaches, which blend the \emph{outputs} of diffusion models and rely on brittle heuristics to define \textit{per-frame} blending masks. By refining the input condition, rather than model outputs, our approach allows the diffusion model to infer how each blocking pose should influence the in-between motion without requiring heuristics for setting per-frame blending masks. The result is the first diffusion model that can robustly convert blocking poses into detailed, natural looking character animations, offering direct support for an important animation workflow.

%% file: relatedwork.tex
\section{Problem and Related Work}

\paragraph{Problem statement} We seek a solution to the motion detailing problem, where 
an animator provides a set of \textit{K} blocking poses $\mathbf{x}^{k} \in \mathbb{R}^{J \times D}$ and positions each pose on a timeline of $F$ frames.
Each pose is represented by a \textit{D}-dimensional feature of $J$ joints. We expect blocking poses to be coarsely specified by the animator, so in a blocking pose only the values for a few important joints (center of mass, hips, shoulders, etc.) may reflect the animator's intent.  The output of the system should be a complete animation $\mathbf{Y} \in \mathbb{R}^{F \times J \times D}$, that specifies joint values for all frames, and yields an animation that is well aligned with gist of the blocking poses and also exhibits natural-looking, detailed motion. For example, if an animator authored a blocking pose $\mathbf{x}^{k}$ at frame $f$, $\mathbf{Y}$ should contain a pose that is a plausible, more detailed version of $\mathbf{x}^{k}$ at a frame near (but not necessarily at) $f$.

\paragraph{\revision{Motion Inbetweening}}\revision{Motion inbetweening focuses on generating smooth, natural transitions given a set of temporally sparse keyframe constraints. Neural network–based methods dominate this task, including RNN-based architectures~\cite{harveypal2,2018-MIG-autoComplete,harvey2020robust}, CNN-based architectures~\cite{Zhou2020GenerativeTL,nemf}, autoencoder-based architectures~\cite{learnedmotionmatching,Kaufmann_2020,oreshkin2022motion,starke,longterm}, and transformer-based architectures~\cite{twostagetransformers, faceinbetween, skelbetween, mo2023continuousintermediatetokenlearning}. 
More recently, diffusion models have emerged as a promising solution for inbetweening due to their ability to generate diverse, high-quality motions~\cite{cohan2024flexible,genmo2025}. 
}

\revision{
Since blocking poses are also temporally sparse, motion inbetweening may seem sufficient for our purposes. However, by definition, inbetweening treats input keyframes as precise specification; they are expected to appear in the output at exactly the same time and with exactly the same pose as in the input. Some approaches even enforce this strictly, either by directly overwriting the output with input keyframes~\cite{wei2023enhanced,tseng2022edge,genmo2025} or by explicitly optimizing the output to match them~\cite{xie2023omnicontrol,pinyoanuntapong2025maskcontrolspatiotemporalcontrolmasked}. However, blocking poses capture the gist of the desired motion, and are not necessarily the final keyframes that should appear in the output. As a result, motion inbetweening methods that strive to perfectly preserve input keyframes produce unnatural results when blocking poses are coarsely or imprecisely specified~\cite{goel2025generative}. }

\paragraph{Retiming Blocking Poses} Goel et al.~\shortcite{goel2025generative} took a step toward support for motion detailing by contributing a diffusion motion inbetweening model (which we refer to as \textit{R} in this paper), that generates high-fidelity motion from keyframes, even when those keyframes (much like blocking poses) are imprecisely timed. Like standard motions diffusion models, \textit{R} is trained to reverse a forward diffusion process; it acts as a denoiser that takes as input a diffusion timestep $t$, a motion $\mathbf{Y}^{R,t}$ that has noise correponding to timestep $t$, and keyframe condition $\mathbf{X} \in \mathbb{R}^{F \times J \times D}$, i.e., $R(\mathbf{X}, t, \mathbf{Y}^{R,t})$. Specifically, in the case of Goel et al.~\shortcite{goel2025generative}, the condition $\mathbf{X}$ is formed by linearly interpolating the input key poses. At inference, during iterative denoising process, the model generates predictions of the final motion that we denote $\tilde{\mathbf{Y}}^{R}$. We point readers to~\cite{ho2020denoising,tevet2023human} for a more thorough discussion on diffusion and its applications to motion generation.

The \textit{R} model is trained to retime $\mathbf{X}$ and synthesize realistic inbetween motion to produce a completed, detailed output $\tilde{\mathbf{Y}}^{R}$.
While \textit{R} is a promising step toward generating motion from blocking poses, it is primarily designed to retime and inbetween precisely posed keyframes. Specifically \textit{R} is trained to preserve the original posing, and thus does not address an important aspect of the motion detailing problem: the process of refining and enriching the blocking poses themselves while generating a natural $\tilde{\mathbf{Y}}^{R}$.

\paragraph{Loose adherence to keyframe constraints} The challenge of both respecting the gist of blocking poses while also adding detail is similar to the goals of inference-time imputation or blending techniques~\cite{tevet2023human,shafir2024human,goel2024iterative} that involve blending the output of each conditioned diffusion denoising inference step with an unconditioned (or differently conditioned~\cite{Avrahami_2022}) generation. These techniques aim to soften the influence of the constraints provided as a condition. For example, one variant blends the intermediate conditioned prediction $\tilde{\mathbf{Y}}^{R} = R(\mathbf{X}, t, \mathbf{Y}^{R, t})$ with $\tilde{\mathbf{Y}}^{U} = U(t, \mathbf{Y}^{U, t})$
according to a blending mask $\mathbf{M} \in \mathbb{R}^{F \times J}$ where \textit{U} is an unconditioned diffusion model. The goal is to soften the influence of constraints provided by the input condition $\mathbf{X}$ (which in the case of motion detailing are coarse blocking poses), with the values of $\mathbf{M}$ set to reflect how much the final output can diverge from the constraint. However, it is often unclear how to set up $\mathbf{M}$. 
For example, masks which assign non-zero weights only to input keyframes and zero elsewhere result in discontinuous motion~\cite{cohan2024flexible}. Smoothing masks in time (e.g., using linear falloffs) can address discontinuities~\cite{shafir2024human}, but relies on brittle heuristics and assumptions about how the influence of constraints should decay over time.

Another approach for implementing loose adherence to keyframes is inference-time reconstruction guidance. This technique modifies~\textit{U}’s predictions at each denoising step to better satisfy given constraints based on a tunable guidance strength parameter controlling how strongly each constraint is enforced. In the context of motion detailing the guidance weight could reflect the animator’s intended tolerance for deviation from the key pose. However, reconstruction guidance alone has been shown to be insufficient for producing natural motion from keyframes~\cite{cohan2024flexible}.

%% file: method.tex
\section{Method}

Our motion detailing solution grew out of failed attempts to construct temporally smooth masks $\mathbf{M}$ for blending blocking-pose-conditioned and unconditional diffusion output. A mask that is too ``narrow'' in time risks discontinuous motion. A too-wide mask risks keeping large segments of the output motion too close to the undetailed blocking input. Further, how much neighboring frames should be influenced by the blocking poses in the final detailed animation might be different per-animation, per-pose, or even per-joint.  

\revision{Our key idea is rather than requiring the animator to make complex judgments about the temporal influence of blocking pose constraints, keeping the condition $\mathbf{X}$ constant throughout the diffusion process, and blending intermediate diffusion outputs, we instead refine $\mathbf{X}$ throughout the diffusion process by blending it with an unconditioned output. This effectively adds detail to the input as it conditions the diffusion model.}
  
\begin{figure}
    \centering
    \includegraphics[page=1,trim={130 130 500 60},clip,width=0.9\linewidth]{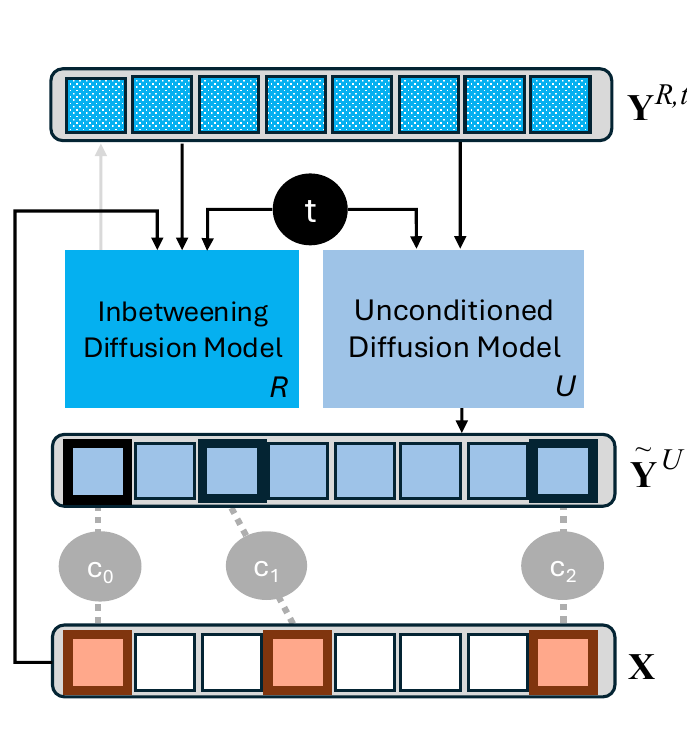}
    \caption{\textbf{Constraint Refinement}. We simultaneously refine the input $\mathbf{X}$ to model \textit{R} while generating $\tilde{\mathbf{Y}}^{R}$. Every N timesteps, we blend input blocking poses with an unconditioned proposal $\tilde{\mathbf{Y}}^{U}$ according to user-provided tolerance weights $\mathbf{C} \in \mathbb{R}^{K \times J}$. The resulting blended input is then used to condition subsequent inference steps of model $\mathit{R}$, and remains fixed until the next blending event. Importantly, blending modifies \textit{input blocking poses} rather than the model output, avoiding the need for dense blending weights.}. 
    \label{fig:constraint_refinement}
\end{figure}

Specifically, every \textit{N} diffusion inference time steps, in addition to generating an intermediate prediction $\tilde{\mathbf{Y}}^{R} = R(\mathbf{X}, t, \mathbf{Y}^{R, t})$, we also take a separate diffusion step on the same $\mathbf{Y}_{R,t}$ using \textit{U}, yielding $\tilde{\mathbf{Y}}^{U} = U(t, \mathbf{Y}^{R, t})$. Then given $\tilde{\mathbf{Y}}^{U}$, we recreate the input condition $\mathbf{X}$ as follows. For each of the $k$ input blocking poses $\mathbf{X}_{f_k}$, we identify $\tilde{\mathbf{Y}}^{U}_{f_k^*}$, where $f_k^*$ is the index of the pose in $\tilde{\mathbf{Y}}^{U}$ that is most similar to $\mathbf{X}_{f_k}$. Since \textit{R} may retime $\mathbf{X}$ during generation, $f_k^*$ need not be the same as $f_k$. In our implementation, we search within $\pm10$ frames of $f_k$ for a closest match. We then refine the pose $\mathbf{X}_{f_k}$ by blending it with the corresponding pose from the unconditioned model, weighted by a per-blocking-pose joint tolerance  vector $\mathbf{C}_k \in \mathbb{R}^{J}$, 

\begin{equation*}
     \mathbf{X}_{f_k} = \mathbf{C}_k \odot \mathbf{X}_{f_k} + (1 - \mathbf{C}_k) \odot \tilde{\mathbf{Y}}^{U}_{f_k^*}
\end{equation*}

\noindent
where $\odot$ denotes elementwise multiplication. We illustrate this in Fig.~\ref{fig:constraint_refinement}. High values in $\mathbf{C}_k$ favor adherence to joint values in the original blocking pose; low values incorporate more of \textit{U}'s proposal.  

Our reasoning is that \textit{U} has a strong prior for plausible kinematic motion. Since $\mathbf{X}_{f_k}$ is a blocking pose, it may initially be underspecified or incomplete. $\tilde{\mathbf{Y}}^{U}$ can be thought of as containing a version of $\mathbf{X}_{f_k}$ that is better aligned with \textit{U}'s learned distribution of detailed motion. Since $\tilde{\mathbf{Y}}^{U}$ is indirectly influenced by $\mathbf{X}$ through its dependence on $\mathbf{Y}^{R, t}$, the resulting blend remains coherent. Following each blending step, we optionally apply IK postprocessing to the refined $\mathbf{X}_{f_k}$ to resolve any ground penetration artifacts, then reconstruct $\mathbf{X}$ by linear interpolation of the modified blocking poses $\mathbf{X}_{f_k}$.

Note that unlike output blending approaches where the blending mask is specified \emph{per-frame}, our method requires only a \emph{per-blocking-pose} joint blending mask $\mathbf{C}_k$. \emph{In other words, the animator only needs to specify how closely the output animation should adhere to input blocking poses at the granularity of the blocking poses themselves}, and it is the responsibility of the diffusion prior to determine how to spread that adherence out over the animation timeline.

%% file: results.tex
\section{Results}

Our goal is to produce realistic, detailed $\mathbf{Y}$ from blocking pose constraints in $\mathbf{X}$. Our evaluation focuses primarily on $\mathbf{Y}$'s realism, and also on $\mathbf{Y}$'s ability to loosely adhere to $\mathbf{X}$. We perform constraint refinement every 100 denoising steps. Further implementation details, e.g., hyperparameters, are included in the Supplement. 
\subsection{Qualitative Evaluation}
\label{sec:qual-eval}

\paragraph{\revision{Choosing and tuning tolerance weights}} \revision{In animating with our tool, we developed the following procedure for setting tolerance parameters. We found that initializing $C_{k,j}$, the tolerance weight for joint \textit{j} in the \textit{k}'th blocking pose, to a default value of 0.85 provides a good balance between adherence to input pose specification and naturalness. When practitioners have strong preferences for preserving a joint’s configuration from a blocking pose, e.g., the hip and knee during a kick, we found that a higher tolerance of $C_{k,j} > 0.95$ ensures the system does not unintentionally ``rewrite'' the meaning of the pose during refinement (e.g., turning a kicking pose into a stepping pose). For unposed joints, e.g., elbows and wrists in the kicking pose, we recommend assigning lower $C_{k,j}$ so the system can adjust these joints more freely. While we recommend these values as good defaults based on our extensive use of the system, we believe practitioners should always have the ability to increase or decrease per-joint adherence as needed to meet creative goals.}

\paragraph{\revision{Discussion.}} We have used our method to transform a diverse range of blocking poses created by an animator into detailed, realistic motion. The animator authored blocking poses using a traditional web-based keyframe animation interface, which we will release alongside our code. In the accompanying videos and Fig.~\ref{fig:banner}, we showcase results from a variety of character motion types, including locomotion, martial arts, and ballistic actions, demonstrating that our method preserves the overall structure of the blocking motion while inferring natural looking, detailed motions. \revision{We discuss a simple extension allowing the technique to handle arbitrary-length motions in the Supplement.}
We also provide videos comparing our method to \textbf{\textit{R}-NoTolerance} and \mbox{\textbf{\textit{R}-SoftMask}} baselines as described in Section~\ref{sec:quant-eval}. These baselines exhibit less detailed motion than our method; \textbf{\textit{R}-NoTolerance} does not refine the input blocking poses, resulting in motion that lacks detail, particularly in unposed joints where the input is underspecified. \textbf{\textit{R}-SoftMask} often leads to overly-smooth and unnatural motion.

\subsection{Quantitative Evaluation}
\label{sec:quant-eval}

We evaluate the impact of blocking pose inputs on generated motion quality. Our goal in this evaluation is to study how different strategies for loosening adherence to blocking poses in $\mathbf{X}$ affect the quality of resulting motion $\mathbf{Y}$.

\textbf{Experimental Setup} Following the evaluation set-up in~\cite{goel2025generative}, we splice all motions in the HumanML3D train and test dataset to 60-frame clips, and train motion diffusion models \textit{R} and \textit{U} on the train set. 

To evaluate how well different methods handle input blocking poses in $\mathbf{X}$, we first require a dataset of such inputs. However, no existing dataset provides blocking-level animations at scale. To address this, we instead construct a synthetic benchmark by simulating blocking $\mathbf{X}$ from realistic full-body motions $\mathbf{Y}$ in the HumanML3D test set.

Our procedure for creating $\mathbf{X}$ is largely inspired by Cooper et al.~\shortcite{gameanim}, which characterizes typical blocking poses. First, blocking animations are often temporally sparse: animators specify a handful of key moments and leave the rest for later refinement. To simulate this, we select up to 10 keyframes per full motion sequence $\mathbf{Y}$ at random and mask out remaining frames.

Second, blocking poses are often spatially incomplete. Blocking poses typically specify the root and a subset of major joints critical for defining the character’s intent—most commonly the shoulders, elbows, hips, and knees~\cite{gameanim}. Joints such as the spine, wrists, ankles, hands, feet, and neck are often left unposed at this stage. To mimic this structure, we define a set of ``important joints'' comprising the root, shoulders, elbows, hips, and knees. For each motion, we retain a random subset of these important joints and set all other joints to a neutral ``unposed'' configuration (identity rotation). We always include the root joint, which, as emphasized by Cooper and others, is almost always defined in a blocking pose.

Finally, blocking poses are often only approximately timed. To account for this, we temporally perturb each selected pose by a small random offset ($\pm$5), simulating the imprecise timing commonly observed in early animation stages. When these poses are placed on the timeline, they form the input sequence $\mathbf{X}$. We repeat this procedure for the entire test set, resulting in a dataset of $\mathbf{X}$s that are temporally sparse, spatially incomplete, and imprecisely timed, mimicking the characteristics of real-world blocking poses.

\textbf{Baselines} We compare our proposed keyframe tolerance approach for detailing $\mathbf{X}$ against several alternatives.
\begin{itemize}
    \item \textbf{Ours (\textit{R}-Tolerance)}: Our constraint refinement procedure on model \textit{R}, using a tolerance of c for all blocking poses. For simplicity we apply the same tolerance for all joints in all blocking poses. We show results for c=0.85 and 0.50.
    \item \textbf{\textit{R}-NoTolerance}: \textit{R}, run without constraint refinement. This is the same as formulation~\cite{goel2025generative}, or can be thought of as running our method with c=1.0.
    \item \textbf{\textit{R}-DiffusionBlending(c)}: Blend the unconditioned proposal with the \textit{outputs} of \textit{R}. We create a length-\textit{F} blending mask $\mathbf{M} \in \mathbb{R}^{F \times J}$ with value tolerance \textit{c} on keyframe indices and 0 elsewhere. This can be thought of as temporal imputation-based inpainting~\cite{tevet2023human,tseng2023edge}.
    \item \textbf{\textit{R}-SoftMask(c)}: Same as above, but apply a linear fall off to all tolerances to create a smoother blending mask from the provided tolerances. This can be thought of as soft temporal inpainting, as proposed in~\cite{shafir2024human}.
    \item \textbf{\textit{U}-Guidance($\mathbf{w}$)} Generate a motion with unconditioned model \textit{U}, while using reconstruction guidance (weight=$\mathbf{w}$) on intermediate generations to match blocking poses in $\mathbf{X}$.
    \item \revision{\textbf{CondMDI}\cite{cohan2024flexible} a recent diffusion-based motion inbetweening model trained to exactly match the posing and timing of input keyframes in $\mathbf{X}$.}
\end{itemize}

\revision{\textbf{CondMDI} is a baseline diffusion-based motion inbetweening model that expects the posing and timing of input keyframes to appear in the output exactly as they appear in the input; this baseline evaluates the ability of an inbetweening model to handle coarse blocking poses. }\mbox{\textbf{\textit{R}-NoTolerance}} evaluates the ability of an inbetweening model that expects input poses to be precise in pose, but potentially imprecise in timing, to handle blocking poses. \textbf{\textit{R}-DiffusionBlending(c)} illustrates the performance of test-time imputation on the \textit{outputs} of the model, rather than the inputs as in our proposed procedure. Informed by prior work~\cite{cohan2024flexible} which shows that sparse temporal imputation often produces jerky, unnatural motion, we include \textbf{\textit{R}-SoftMasking(c)}, which turns sparse blocking pose tolerances into dense values using via linear falloffs (as proposed by~\cite{shafir2024human}). \mbox{\textbf{\textit{U}-Guidance(w)}} tests whether guidance scale alone can control how strictly $\textit{U}$ matches input poses and generate realistic motion. 

\begin{table}[t!]
\centering
\resizebox{\columnwidth}{!}{%
\begin{tabular}{l c c c c}
\hline
\textbf{Metric} & FootSkate $\downarrow$ ($10^{-3}$) & Jitter $\downarrow$ ($10^{-2}$) & FID $\downarrow$ & KE $\downarrow$ ($10^{-2}$) \\
\hline
Ours (c=0.85)             & 5.96 & 0.241 & 0.058 & \revision{\textbf{0.105}} \\
Ours (c=0.50)            & \textbf{5.14} & \textbf{0.223} & \textbf{0.038} & \revision{0.106} \\
\hline
\textit{U}-Guidance (w=5.0) & 9.86 & 2.562 & 0.118 & \revision{11.455} \\
\textit{U}-Guidance (w=1.0)  & 7.21 & 1.274 & 0.043 & \revision{1.101} \\
\hline
\textit{R}-DiffusionBlending (c=0.85)   & 9.62 & 7.783 & 0.211 & \revision{0.684}  \\
\textit{R}-DiffusionBlending (c=0.50)   & 10.36 & 6.021 & 0.146 & \revision{1.163}  \\
\textit{R}-SoftMask (c=0.85)   & 6.70 & 0.321 & 0.063 & \revision{0.130} \\
\textit{R}-SoftMask (c=0.50)   & 5.97 & 0.284 & 0.040 & \revision{0.144} \\
\hline
CondMDI & 29.51 & 2.293 & 0.305 & \revision{0.664} \\
\hline
\textit{R}-NoTolerance       & 8.19 & 0.282 & 0.077 & \revision{0.145} \\
\hline
\end{tabular}
}
\caption{Our method achieves the best overall motion quality.}
\label{tab:tolerance_results}
\end{table}

\textbf{Metrics.} We evaluate motion quality using \textit{Jitter}, \textit{FootSkate}, and \textit{FID} metrics; please see our Supplemental for details. We also report keyframe error (KE), the L2 error between local-space joint positions of input blocking poses and corresponding frames in generated motion. While perfect KE is not the goal, especially given the incomplete nature of input poses, this metric simply serves as a rough indicator of fidelity to the input. Results are in Table~\ref{tab:tolerance_results}.

\textbf{Discussion.} Our method achieves the highest overall motion quality, producing the lowest jitter, foot-skate, and FID among all evaluated approaches. We see that motion quality metrics remain relatively stable across different values of \textit{c}. As expected, KE scores get smaller as \textit{c} grows; higher confidence in input blocking poses indicate to the model that they should be more tightly adhered to. \revision{Our model also achieves the best KE score, meaning it generates the most realistic motion while also adhering most closely to the overall structure of the input blocking poses. }

\revision{\textbf{\textit{R}-NoTolerance} has a low KE, but the result of trying to adhere to imprecise, rough blocking poses is diminished motion quality, evidenced by higher Jitter, FID, and FootSkate scores. Trained to expect input keyframes to be precisely posed and timed, \textbf{CondMDI} also struggles when faced with coarsely specified input and produces low-quality motion overall. Both approaches may deviate from input blocking poses (as reflected in higher KE), but, without an interface like tolerance weights to relax constraint influence, they do so in an uncontrollable way that can yield unnatural results.}

Another important set of baselines are \mbox{\textbf{\textit{R}-SoftMask(c)}} and \textbf{\textit{R}-DiffusionBlending(c)}, which blend the \textit{outputs} of \textit{R} with an unconditioned proposal according to a blending mask. Similar to prior work, we find that the temporally sparse blending masks in \textbf{\textit{R}-DiffusionBlending(c)} produce noticeable discontinuities before and after each keyframe, reflected in high jitter score. \mbox{\textbf{\textit{R}-SoftMask(c)}} uses a linear fall-off around each blocking pose tolerance, resulting in a temporally smoother blending mask. However, this approach makes assumptions about the underlying motion and often results in unnatural, overly smoothed motion and uncompetitive performance on motion quality metrics. Finally, guidance-based baselines yield poor motion quality, exhibiting high levels of foot skating and elevated FID scores. They do not preserve general structure of input blocking poses as well and produce less natural motion.

\subsubsection{\revision{Choice of \textit{N}}} 
\begin{figure}
    \centering
    \includegraphics[width=\linewidth]{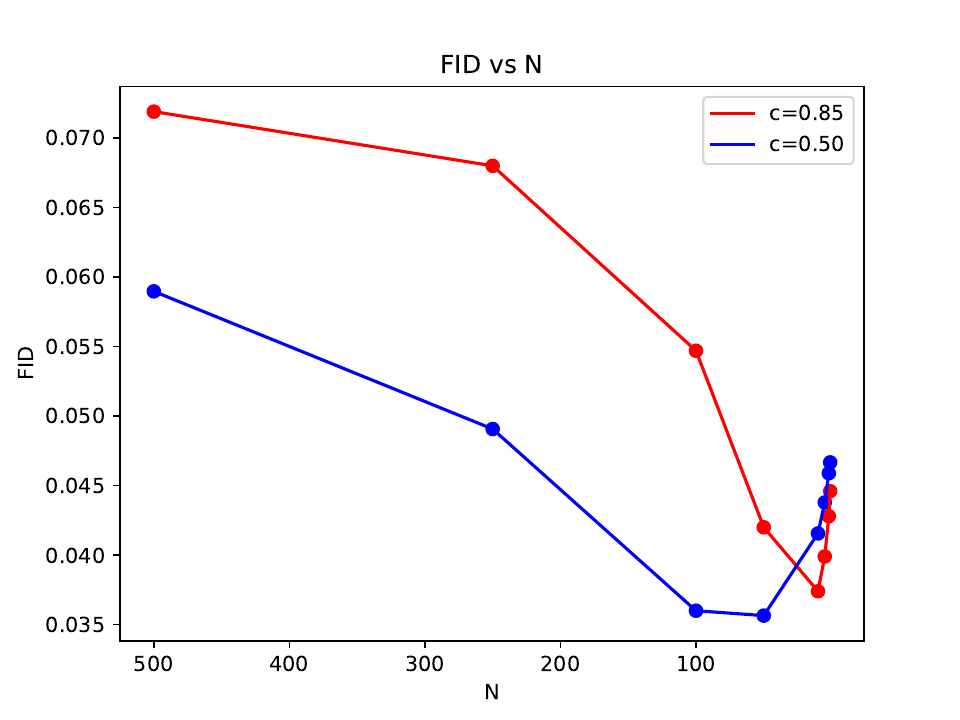}
    \caption{\revision{Constraint refinement is applied every $N$ denoising steps. Notice how FID 
    improves as $N$ gets smaller (meaning more refinement steps), until a turning point between after which FID worsens.}}
    \label{fig:nvsfid}
\end{figure}

\revision{Earlier, we proposed applying our constraint refinement every $N=100$ denoising steps. To evaluate the effect of different values of \textit{N} on the generated motion quality, we use the experimental setup from Section 4.2, e.g., synthesizing input blocking poses $\mathbf{X}$ from realistic motions $\mathbf{Y}$ in the test set. We run our proposed method on the dataset with different values of $N$, then evaluate FID on generated motions. We report results for $c=0.85$ and $c=0.5$ over several candidate $N$, ranging from $1$ to $500$. For this ablation, we omit IK cleanup of $\mathbf{X}$, as it would be prohibitively slow for small $N$. As shown in Figure~\ref{fig:nvsfid}, quality improves as N decreases, up to a turning point beyond which FID worsens. }

\revision{We attribute this drop to the diffusion model having too few denoising steps between constraint refinements to produce as high-quality predictions. Additionally, smaller $N$ values increase computational cost, as each refinement step requires an additional forward pass of $U$. We chose $N=100$ as a good balance of cost and maintaining good quality scores.}

%% file: conclusion.tex
\section{Conclusion}
\revision{\textbf{Limitations and future work.} While our method can generate natural motion from blocking-level input poses, the generated motions can suffer from common artifacts like footskate. These could potentially be addressed by IK-based postprocessing, additional training-time losses, or physics-based tracking~\cite{yuan2023physdiff}. 
Our method fails to detail blocking poses in cases where they lack essential information about or structure of the desired motion.
For example, if the animator wants to create a jumping motion, providing only two T-poses on the ground corresponding to the start and end of the motion is unlikely to yield a detailed jumping motion. The animator would need to provide an additional blocking pose of the character in the air for the model to infer the jump intent. An interesting direction for future work, then, would be to incorporate additional conditioning to the system, like text prompts or video demonstration, as a way to clarify intent~\cite{maneesh:blackboxes:2023} when blocking inputs are too ambiguous. }

We presented a simple and effective inference-time technique to instrument a retiming inbetweening model with the ability to produce detailed motion from sparse blocking poses.
This work takes a step toward aligning modern diffusion-based synthesis tools with real-world animation workflows. We hope that continued progress in controllable generative models, grounded in workflows inspired by animator practice, will bring us closer to intelligent animation systems that truly support animators.

\begin{acks}
Purvi Goel is supported by a Stanford Interdisciplinary Graduate
Fellowship.  We thank the anonymous reviewers for constructive feedback; Maneesh Agrawala, Vishnu Sarukkai, Sarah Jobalia, Sofia Di Toro Wyetzner, and Zander Majercik for helpful discussions.
\end{acks}